\documentclass[aps,prb,twocolumn]{revtex4}
\usepackage{graphicx,amsmath,amssymb,bm}
\begin{document}
%\begin{document}
	
\title{Superfluid Density in Conventional Superconductors: From Clean to Strongly Disordered}
\author{Surajit Dutta$^1$, Pratap Raychaudhuri$^1$, Sudhansu S. Mandal$^2$, T.V.Ramakrishnan$^3$}

\affiliation{$^1$Tata Institute of Fundamental Research, Mumbai 400005, India  \\
	$^2$Department of Physics, Indian Institute of Technology, Kharagpur 721302, India\\
	$^3$Department of Physics, Indian Institute of Science, Bangalore 560012, India}

\begin{abstract}

The highly convergent form of superfluid density in disordered conventional superconductors available 
 in the literature 
 and independently obtained by us following the approach of an earlier paper [Phys. Rev. B $\bm{102}$, 024514 (2020)] 
 has been reformulated to separate out the generally used so-called `dirty-limit' 
 term and an additional term. We use this new  expression for making an extensive comparison with previously published experimental data and show that the former, generally used, term is {\em not} sufficient for analyzing these results. We point out that consequently, there is a large regime (disordered superconductors with moderate to no disorder) where theoretical predictions need to be confronted with experiment. 
\end{abstract}

\maketitle

\section{Introduction}

The additional free energy ${\cal F}$ of a superconductor depends on its nonzero
 superfluid velocity ${\rm \bf v}_s$ as ${\cal F} \sim (\rho_s/2) \int d\bm{r}\, {\rm \bf v}_s^2$ where $\rho_s$ is the superfluid stiffness or phase rigidity, analogous to the mass (see e.g. Ref.\cite{Coleman}). This gauge invariant superfluid velocity ${\rm \bf v}_s$ is related to the phase $\theta$ of the superconducting order parameter as  ${\rm \bf v}_s = (1/m_e)(\bm{\nabla}\theta - 2e\bm{A})$; here  $\bm{A}$ is the vector potential, $e$ and $m_e$ are the charge and mass of an electron respectively, and we set $\hbar =1$. Experimentally, one measures the magnetic penetration depth $\lambda$ which is related to the superfluid density $n_s$ as\cite{Tinkham} $\lambda^{-2} =\mu_0e^2n_s/m_e$. The superfluid density $n_s$ is proportional to the superfluid stiffness; $n_s = (4/m_e)\rho_s$. We use the above relation between the experimentally measured penetration depth $\lambda$  and the calculated $\rho_s$ to compare in detail theoretical results with experiment, and suggest that there is a large regime of disorder in relatively clean systems so that measurements are needed here, to also establish the clean London limiting value.   

The solely diamagnetic response of the electron system to an external magnetic field leads to 
 $n_s^d = n$, the electron density. This is the London value which also follows for the ground state ($T=0$) from Galilean invariance, for a homogeneous continuum.  However, the actual superfluid density is less than $n_s^d$ due to the paramagnetic response of the system: $n_s = n_s^d - n_s^p$, $n_s^p$ being the paramagnetic contribution to the superfluid density. For a clean conventional Bardeen-Cooper-Schrieffer (BCS) superconductor \cite{BCS}, $n_s^p =0$ at zero temperature and is exponentially small at low temperatures because of the presence of the quasiparticle gap. However, $n_s^p$ grows with temperature and eventually becomes equal to $n_s^d$ at the superconducting critical temperature $T_c$ where $n_s$ vanishes. In disordered superconductors, $n_s^p \neq 0$ at zero temperature ($T=0$), and the resulting superfluid density is disorder dependent and is smaller \cite{A-G} than the London limiting value at $T=0$. This, and the temperature dependence of $n_s$ have been discussed in literature \cite{AG2,AGD,Weiss,Nam,Kogan,M-R}.

The effect of static, short range nonmagnetic disorder on superconductors is most simply characterized 
 by a broadening $\Gamma \ll \epsilon_{\rm F}$ of the electron spectral density (here $\epsilon_{\rm F}$ is the Fermi energy \cite{Coleman,A-G}. Microscopic calculations generally use on site or zero range disorder with a Gaussian probability distribution of its strength related to this broadening. The effect of disorder on electrons is mostly implemented in the Born approximation, where it leads to a finite lifetime $\tau = (1/\Gamma)$ of electronic states. Such a treatment neglects Anderson localization effects\cite{Ma-Lee}. In this approximation, it is well known that in the so called `dirty limit', i.e. for $\Delta_0/\Gamma <<1$, $n_s$ at $T=0$ scales \cite{A-G} with the dc conductivity $\sigma = ne^2\tau/m_e$ in the normal state, i.e., $n_s(T=0) = \sigma (\pi m_e\Delta_0/ e^2) = n\pi\Delta_0\tau$, where $\sigma$ is the electrical conductivity of the system, $n$ is the normal electron density, and $\Delta_0 $ is the gap at $T=0$. We note that $\Delta_0$ is independent of disorder, according to Anderson's theorem\cite{Anderson}. A  generalized form of this zero-temperature superfluid density at finite temperatures, namely
\begin{equation}
n_s(T) = n\pi \tau \Delta(T)\,\tanh \left(\frac{\Delta (T)}{2k_BT} \right) \,
\label{Eq-0}
\end{equation}
 is often used for analyzing experimental data \cite{Lemberger07,Mondal11,Mandal20}; 
 where $\Delta(T)$ is the gap at the temperature $T$. However, this expression has also been derived \cite{AG2,Kogan} in the dirty limit. Clearly, $n_s(T)$ in Eq.~(\ref{Eq-0}) cannot be valid for all $\tau$ because for  $\tau$ large enough such that $\Delta_0 \tau > 1/\pi$, the superfluid density $n_s (T=0)$ exceeds the maximum possible London limiting value $n$.

In this paper, we exhibit the superfluid density as a sum of the commonly used  
 term (\ref{Eq-0}) and another term in the following way. We reformulate an expression (\ref{Eq-6}) of superfluid density \cite{AGD, Nam, Weiss, Scheffler}, which is a convergent sum of Matsubara frequencies only 
 %( Sudhansu: would you like to explain here, maybe in a footnote, how and where thuese terms occur in these papers; e.g. AGD does not seem to have the Matsubara sum form explicitly, but inferred from the form they have; Thiemann quote it from Nam in their supplementary material)
 and which shows explicitly that $n_s$ vanishes when $\Delta$ vanishes. This frequency sum is converted into a contour integral over complex frequencies, and displays two simple poles at $\pm \Delta$ and branch cuts for the domains $(\Delta,\infty)$ and $(-\infty,-\Delta)$. The residue of the simple poles provides the contribution (\ref{Eq-0}) generally used for the analysis of experimental data.
 %, and referred to here as the phenomenological term. 
 We have derived an additional contribution arising from the branch cuts; this competes with the former as they are opposite in sign. We find that the contribution of the latter is insignificant if $\Delta_0\tau \lesssim  10^{-3}$; it begins to be relevant for  $\Delta_0\tau \sim  5\times 10^{-3}$. Both the contributions increase with  $\Delta_0\tau$, and their difference asymptotically approaches the London limit at $T=0$  
 for $\Delta_0\tau \rightarrow \infty$. The contribution of the latter to superfluid density and thus to the measured absolute value of the penetration depth provides a large regime, which is yet unexplored, for experimental studies of disorder dependent superfluid density in relatively clean superconductors over a wide span in $\Delta_0 \tau$, namely roughly from $10^{-3}$ to $10$, i.e., from the dirty limit to the clean limit.

We also find that temperature dependence of the scaled superfluid density $n_s(T)/n_s(0)$ is 
 almost independent of disorder; this scaled density function is easily obtained in the dirty limit as well as in the pure limit, and is the same. This fact has led to the belief that the dirty limit expression is appropriate for all disorder, including very weak disorder.

Our finding suggests a disorder dependent study with measurement of the {\em absolute} value of 
 the superfluid density as a function of disorder, and provides explicit expressions for it at different temperatures and for different values of disorder.  Unfortunately, not much data is available in the literature where absolute measurement of $n_s$ has been performed, so that our results cannot be easily compared with experiment. In Section III, we analyze some of the available experimental data in superconductors like Nb-doped SrTiO$_{_3}$, Pb, Sn, Nb, NbN, and a-MoGe. The data for $T_c$ and $n$ have been obtained via transport measurements, and the dimensionless parameter $\delta = \Delta_0/(2k_BT_c)$ is obtained from the measurement of $\Delta_0$ in tunneling experiments. We then have just one free parameter $\Delta_0\tau$ which we extract by fitting the above mentioned theoretical expression where we have explicitly shown also the contributions of both the terms in the expressions separately. The extracted values of $\Delta_0\tau$ range from about $5\times 10^{-5}$ to $0.5$. The ratio $\eta$ of the two contributions to $n_{s}(T)$ mentioned above, is almost negligible for {\it a}-MoGe and NbN for which $\Delta_0\tau$ is very small, but it becomes recognizable for the  Nb sample, it becomes more prominent for Pb and Sn, and for Nb-doped SrTiO$_3$ it is the largest amongst all the ones analyzed here.

Section IV is devoted to the outlook and discussion where we have pointed out that many more 
 experiments are needed to be confronted with theoretical prediction as the highest value of $n_s/n$ that has been found in the earlier experiments is about $0.56$, whereas it can go up to $1.0$ for the pure limit that may be attained for the samples with $\Delta_0\tau \sim 10$. We also discuss here the physics that cannot be revealed from the theoretical prediction above.

In appendix A, we have estimated the superfluid density by utilizing the oscillator sum rule for the 
 real part of optical conductivity. We show that it reproduces the clean limit exactly and the dirty limit up to a numerical factor of order unity.

\section{Reformulated Superfluid Density}

A highly convergent expression \cite{AGD} of super-fluid density (see also Refs.\onlinecite{Nam,Weiss,AG2,Kogan}) at finite temperatures for all disorder (excluding the localization regime) is given by
 \begin{equation}
 n_s(T) = \frac{n\pi}{\beta} \sum_{\omega_m}\left[\frac{\tilde{\Delta}^2}{(\tilde{\Delta}^2+\tilde{\omega}_m^2)^{3/2}}\right]
 \label{Eq-6}
 \end{equation}
which is obtained also by a series of successive integration by parts for removing divergences in the approach of Ref.\onlinecite{M-R},
where renormalized frequency, $\tilde{\omega}_m$, and gap, $\tilde{\Delta}$, in terms of Matsubara frequency $\omega_m = \pi (2m+1)/\beta$ and the superconducting gap $\Delta$ can be expressed as 
 \begin{equation}
\frac{\tilde{\omega}_m}{\omega_m} = \frac{\tilde{\Delta}}{\Delta} =1 + \frac{1}{2\tau \sqrt{\Delta^2 +\omega_m^2}}.
		\end{equation}
Here $\beta = 1/(k_BT)$ and the introduction \cite{A-G} of finite electronic life-time $\tau$ in the theory of disordered superconductors through the Nambu-Green's functions for Bogoliubov quasiparticles.  
The superfluid density is explicitly seen to vanish (\ref{Eq-6}) in the absence of isotropic superconducting gap. 	
% It is seen explicitly to vanish when there is no superconducting gap, i.e. when the ${\text{\bf k}}$-independent gap  $\tilde{\Delta}$ vanishes.
The expression (\ref{Eq-6}) of $n_s(T)$ is applicable to both two and three dimensional superconductors provided the localization effect of disorder does not set in for very strong disorder.
However, the expression (\ref{Eq-6}) has not been frequently used for analyzing experimental data because it involves complicated sum over the Matsubara frequency. Here we reformulate Eq.\ref{Eq-6} below in terms of a simple term and a simple integral, and analyze 
available data of the absolute measurements of superfluid density in the next section. 

\begin{figure}[h]
	\includegraphics[scale=0.5]{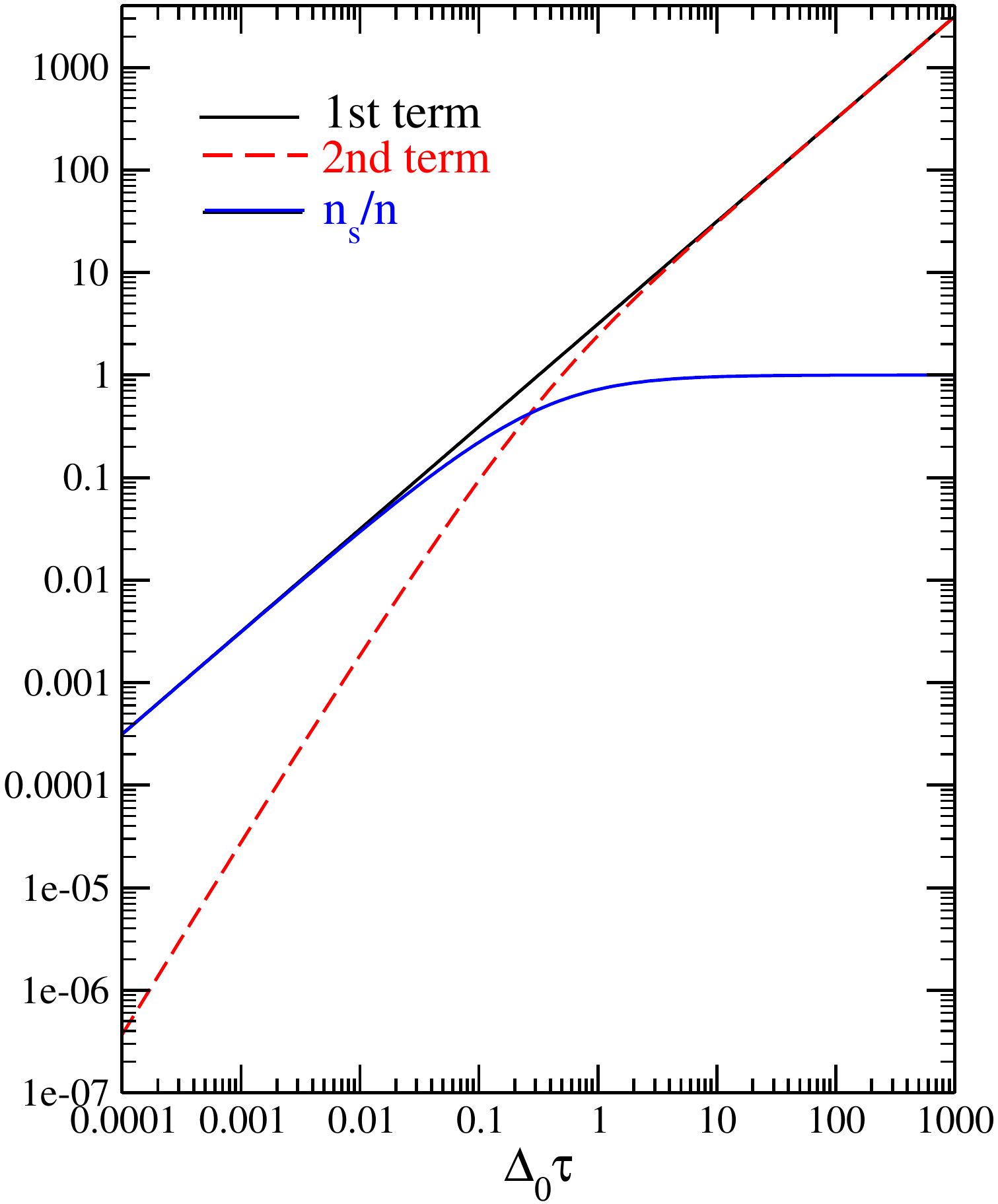}
	\caption{(Color online) Zero temperature contributions for the expression (\ref{Eq-8}) of normalized superfluid density, $n_s/n$, as a function of dimensionless disorder $\Delta_0\tau$: We show the first term, the second term, and their difference  for several decades of $\Delta_0 \tau$. The  nonzero value of the  difference is apparent on this log-log scale in the separate curve for it.}
	\label{Fig:t0}
\end{figure}

The frequency sum in Eq.~\ref{Eq-6} is evaluated in the usual way: we change it into a contour integration for complex $z$ such that the contour contains only the poles at $z=i \omega_m$,
 \begin{equation}
 n_s (T) =n\pi \int_C \frac{dz}{2\pi i} \frac{\Delta^2}{(\Delta^2-z^2)(\sqrt{\Delta^2-z^2}+ \frac{1}{2\tau})} \frac{1}{e^{\beta z}+1}.
 \label{Eq-7}
 \end{equation}
 We now deform the contour to exclude the non-analyticities on the real axis (energy $\epsilon$), namely the simple poles at $z=\pm \Delta$ as well as the branch cut from  $z = \Delta \to \infty$ and from $-\Delta \to -\infty$ (arising from the square root term) so that 
 \begin{eqnarray}
 \frac{n_s}{n} &=& \pi \Delta \tau \tanh \left( \frac{\beta \Delta}{2} \right) \nonumber \\
 & -& \Delta^2 \int_\Delta^\infty d\epsilon \frac{\tanh \left( \frac{\beta \epsilon}{2} \right)}{\sqrt{\epsilon^2-\Delta^2} (\epsilon^2 -\Delta^2 + \frac{1}{4\tau^2})} \, . 
 \label{Eq-8} 
 \end{eqnarray}
 The first term is due to the contribution from the residues of the poles and the second term is due to the branch cut. In the dirty limit $(\Delta_0\tau <<1)$ the latter is much smaller than the first and can therefore be neglected; the contribution of the corresponding branch cut to the superfluid density is negligible. This fact leads to a considerable simplification of calculations in the dirty limit. We note that while the first term in Eq.(\ref{Eq-8}) is present in the superfluid density expression in Ref.\onlinecite{M-R} as well, the second term differs. This finer difference makes the expression (\ref{Eq-8}) consistent in temperature dependence at all disorder.

\begin{figure}[h]
	\includegraphics[scale=0.38]{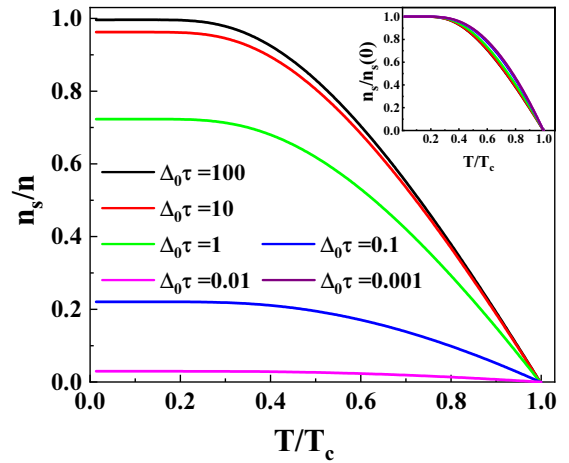}
	\caption{(Color online) Temperature dependence of $n_s$ scaled with electron density for different levels of disorder: $\Delta_0 \tau=10^{-3},\,10^{-2},\,10^{-1},\,1,\,10,\,10^2$ (in the unit of $\hbar$). Temperature is scaled with the BCS $T_c$. Inset: $n_s(T)$ is scaled with $n_{s0} = n_s(T=0)$. Temperature variation of $n_s/n_{s0}$ is almost independent of disorder, although $n_{s0}$ is strongly disorder dependent.}  
	\label{Fig2}	
\end{figure}

 The zero temperature limit of Eq.~(\ref{Eq-8}) yields
 \begin{eqnarray}
 \frac{n_s}{n}(T=0) = \pi \Delta_0\tau -&&\nonumber \\
   \left\{\begin{array}{ll} \frac{(2\Delta_0\tau)^2}{\sqrt{(2\Delta_0\tau)^2 -1}} \tan^{-1} \left( \sqrt{(2\Delta_0\tau)^2 -1} \right) & \text{for} \,\,\,\, 2\Delta_0\tau >1 \\
 \frac{(2\Delta_0\tau)^2}{\sqrt{1-(2\Delta_0\tau)^2} } \tanh^{-1} \left( \sqrt{1-(2\Delta_0\tau)^2}  \right) & \text{for} \,\,\,\, 2\Delta_0\tau \leq 1	
 \end{array}\right. &&\nonumber \\
 && \label{Eq-10a}
 \end{eqnarray}
 Though superficially different from the well known $T=0$ 
 result \cite{A-G,Nam,Kogan,M-R}, 
 this has also the right clean and dirty limits, namely  $n$ and $n\pi \Delta_0\tau$. Although the first term in Eq.~(\ref{Eq-8}) is sufficient for extreme dirty limit $(\Delta_0\tau <<1)$ as mentioned above, 
 it alone is incomplete when $\Delta_0\tau \sim 1$ as it can exceed the London limit, namely the electron density $n$! 
 We show variations of the first and second terms of Eq.(\ref{Eq-8}) and their difference, i.e., $n_s/n$ over several decades of $\Delta_0\tau $ at $T=0$ in Fig.~\ref{Fig:t0}. Contribution of the second term is negligible as it is less by 3 orders of magnitude than the first term when $\Delta_0 \tau = 10^{-4}$. However, the role of the former begins to be significant even for $\Delta_0 \tau = 5\times 10^{-3}$ when the latter is about $3$\% of the former. While both the terms increase with $\Delta_0\tau$, the difference between them asymptotically becomes unity, namely it approaches the disorder-free London limit.The zero temperature value of $n_s$ depends strongly on $\Delta_0\tau$ and attains the pure limit for $\Delta_0\tau \sim 10$ while it has the dirty limit value for $\Delta_0\tau \lesssim 0.005$.

  The temperature dependence of $n_s$ is  numerically calculated using a dimensionless form of the variables and parameters of Eq.~(\ref{Eq-8}) and reinstating $\hbar$ as appropriate:  
\begin{eqnarray}
\frac{n_s}{n} &=& \pi \tilde{\Delta} \left(\frac{\Delta_0\tau}{\hbar}\right) \tanh \left( \frac{\delta\tilde{\Delta}}{T/T_c} \right) \nonumber \\
& -& \tilde{\Delta}^2 \int_{\tilde{\Delta}}^\infty d\tilde{\epsilon} \frac{\tanh \left( \frac{\delta \tilde{\epsilon}}{T/T_c} \right)}{\sqrt{\tilde{\epsilon}^2-\tilde{\Delta}^2} (\tilde{\epsilon}^2 -\tilde{\Delta}^2 + \frac{1}{4(\Delta_0\tau/\hbar)^2})}. 
\label{Eq-10} 
\end{eqnarray}
where $\tilde{\Delta} = \Delta/\Delta_0$, $\tilde{\epsilon} = \epsilon/\Delta_0$, and $\delta = \Delta_0/(2k_BT_c)$. Figure 2 shows the temperature dependence of $n_s/n$ for a wide range of $\Delta_0\tau$ (in the unit of $\hbar$) and using the BCS value of $\delta = 0.882$. %(in units of T_{c}).  
As expected, temperature dependence of $n_s$ at low temperatures is exponentially weak due to the presence of gap $\Delta_0$, but it strongly depends on $T$ beyond a threshold value $T_{\rm th}$ and eventually vanishes at $T=T_c$. Inset of Fig.~\ref{Fig2} shows temperature dependence of scaled $n_s(T)$ by its zero-temperature value $n_{s0}$ for wide range of $\Delta_0\tau$ (scaled by $T_{c}$). The unrecognizable differences of $n_s(T)/n_{s}(0)$ with disorder indicates that the experimental techniques in which absolute value (in lieu of relative value with respect to zero temperature) of $n_s(T)$ is measured is the only one suitable for studying the disorder dependence of superfluid density.

%We also find from Eq.(\ref{Eq-6}) that as expected (and as described at length in Appendix A of the paper), in the disorder free or the clean limit, the superfluid density for all temperatures has the BCS form      
%\begin{equation} 
%\frac{n_s}{n} (\tau \to \infty) =1 +2 \int_{\Delta}^{\infty}   \frac{\partial }{\partial E}\left(  \frac{1}{e^{\beta E}+1}  \right) \frac{E}{\sqrt{E^2-\Delta^2}}dE
%\label{Eq-12}
%\end{equation}
%as we find, for example, in Ref.\onlinecite{Tinkham}. %Here $E= (\xi^2 +\Delta)^{1/2}$.

\section{Comparison with Experiment}
 
\begin{widetext} 
	
  \begin{table}[h]
  	\caption{Experimental data of $T_c$, $\Delta (0)$, $\lambda (0)$, $n$, and normal-state resistivity $\rho_{_N}$, mean free path $\ell$ and effective mas $m^\ast$ of an electron obtained from a number of experiments \cite{Devlin,Megerle,Wyckoff,Kittel,Rinderer83,Lemberger07,Sutton,Karim,Mondal11,Chand,Mattheiss,Chockalingam,Mandal20,Dressel,Lin} in various samples. }
  	\begin{tabular}{|l|l|l|l|l|l|l|l|}\hline
  		Sample &  $T_c$  & $\Delta(0)$  & $\lambda (0)$ & $n$ & $\rho_{_N}$ &$\ell$ & $m^\ast/m_e$\\
  		& (K) & (meV) & (nm) &  ($10^{28}$ m$^{-3}$) & ($\mu\Omega$-m) & $(A^o)$ & \\\hline 
  		Sn & 3.72 \cite{Devlin} & 0.555 \cite{Megerle} & 42.5 \cite{Devlin}& 14.8 \cite{Wyckoff}&  *** & *** &  1.26 \cite{Kittel}\\ \hline
  		Pb & 7.2 \cite{Rinderer83}& 1.34 \cite{Megerle}& 52.5 \cite{Rinderer83}& 13.2 \cite{Wyckoff}& *** & *** & 1.97 \cite{Kittel}\\ \hline
  		Nb (15.3nm) & 8.17  \cite{Lemberger07}&1.525 \cite{Sutton}& 135.08 \cite{Lemberger07}& 5.56 \cite{Wyckoff}& 0.135 \cite{Lemberger07} & 64.6 & 1.81 \cite{Karim}\\  \hline
  		NbN-1\footnote{$n$ and $\rho_{_N}$ of NbN is obtained by interpolation using given data set of Ref.~\onlinecite{Chand}. The three samples correspond to different levels of disorder.} &14.3 \cite{Mondal11}&2.5 \cite{Chand}& 358.4 \cite{Mondal11}&16.85 \cite{Chand} &1.14 \cite{Chand} & 3.65 & 1.0 \cite{Mattheiss,Chockalingam}\\ \hline
  		NbN-2$^{{\rm a}}$& 9.94 \cite{Mondal11}& 1.736 \cite{Chand}& 583.9 \cite{Mondal11}& 11.6 \cite{Chand}& 2.22 \cite{Chand}& 2.41 & 1.0 \cite{Mattheiss,Chockalingam}\\ \hline
  		NbN-3$^{{\rm a}}$ & 8.5 \cite{Mondal11} & 1.485 \cite{Chand}& 759.1 \cite{Mondal11}& 11.76 \cite{Chand}& 2.41 \cite{Chand}& 2.2 & 1.0 \cite{Mattheiss,Chockalingam}\\ \hline
  		MoGe-1 (21 nm)\footnote{Three amorphous MoGe thin films with different thickness (within bracket). The carrier density is measured from Hall effect for MoGe-1 and assumed to remain same for other thickness.} & 7.56 \cite{Mandal20}& 1.28 \cite{Mandal20}& 528 \cite{Mandal20}& 46 \cite{Mandal20}& 1.5 \cite{Mandal20}&1.42 & 1.0 \\  \hline
  		MoGe-2 (11nm)$^{{\rm b}}$& 6.62 \cite{Mandal20}&1.25 \cite{Mandal20}& 554.6 \cite{Mandal20}& 46 \cite{Mandal20}& 1.64 \cite{Mandal20} & 1.3 & 1.0\\ \hline
  		MoGe-3 (4.5nm)$^{{\rm b}}$ & 4.8 \cite{Mandal20}& 1.12 \cite{Mandal20}& 613.07 \cite{Mandal20}& 46 \cite{Mandal20} & 1.44 \cite{Mandal20} & 1.48 & 1.0\\ \hline
  		Nb-doped STO & 0.346 \cite{Scheffler}& 0.052 \cite{Scheffler}& 1349.5 \cite{Scheffler}& 0.011 \cite{Scheffler} & 0.52 \cite{Scheffler} & *** & 4.0 \cite{Lin}\\
  		\hline
  	\end{tabular}
  	\label{Table1}
  \end{table}	 
  
  \end{widetext}
  
  \begin{figure}[h]
  	%\hspace{-2.3cm}
  	\includegraphics[scale=0.3]{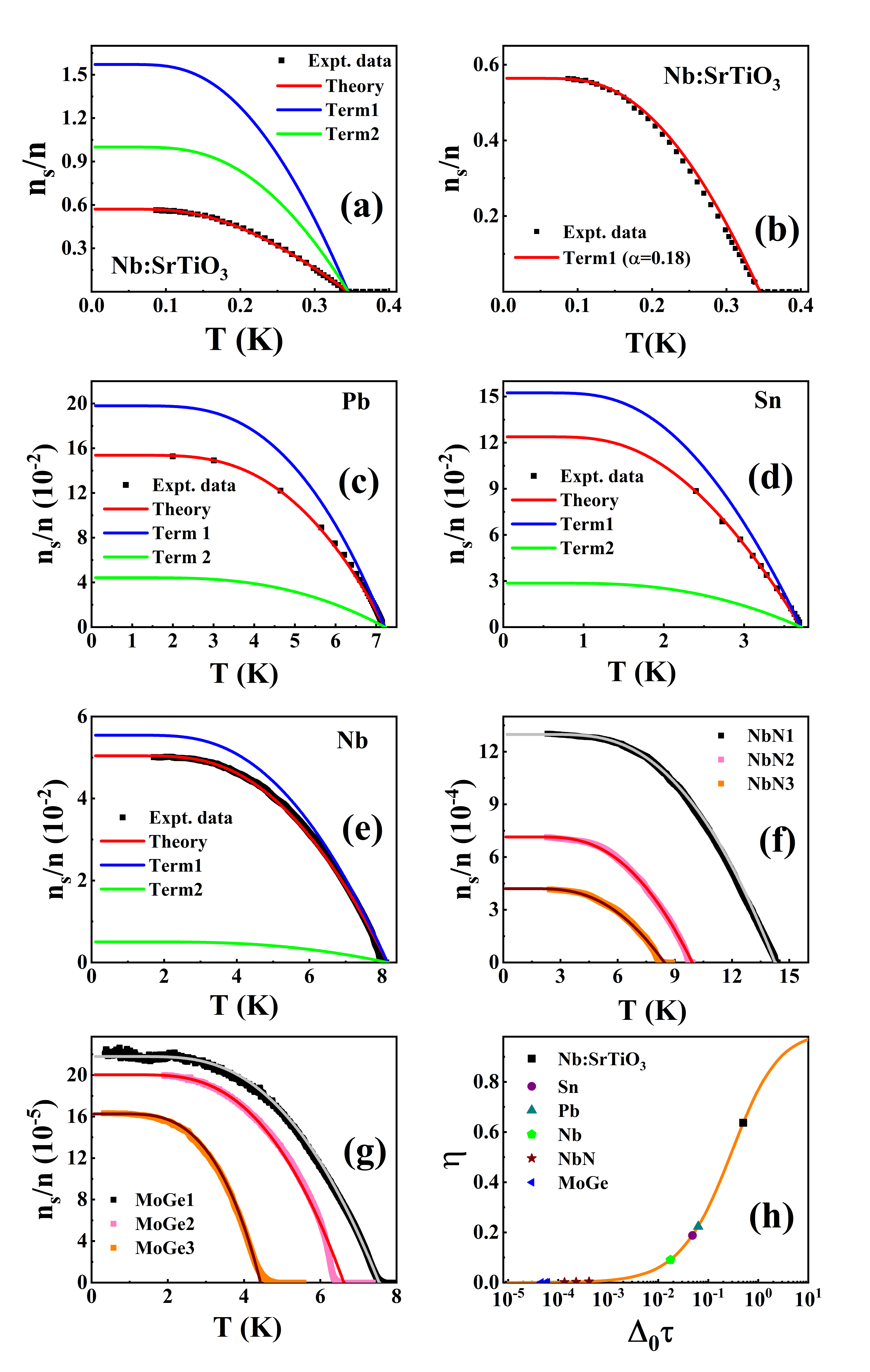}
  	\caption{(Color online) (a) Experimental data (black dots) of $n_s/n$ vs. $T$ for Nb-doped SrTiO$_3$ fitted with Eq.(7); blue and green curves respectively represent the contributions of 1st  and 2nd  terms of Eq. (7). (b) Fit of the same data but with the dirty limit BCS formula which is equivalent to taking only the first term of Eq.(7); the fit deviates at intermediate temperatures. (c)--(e) Experimental data for Pb, Sn crystal, and 15.3-nm thick Nb film respectively; red solid curves are the theoretical fits using Eq.(7). (f) and (g) respectively correspond to temperature dependence of $n_s/n$ for NbN and MoGe films with various thickness; solid lines are the theoretical fits using Eq.(7); this fit is indistinguishable from the contribution of the 1st term of Eq. 71) alone. (h) The ratio of the contributions of the 2nd  and 1st  terms of Eq. (7), $\eta$, vs. the parameter $\Delta_0 \tau$ (in the unit of $\hbar$) (solid line) and the same extracted from the fits mentioned above for various samples (dots).}
  	\label{Fig3}
  \end{figure}

  In this section, we analyze some of the published experimental data of $n_s(T)$ which are extracted from the measured penetration depth using London's formula \cite{Tinkham}:
  \begin{equation}
  n_s = \frac{m^\ast}{\mu_0 e^2} \lambda^{-2} = 2.82 \times 10^{13} \left( \frac{m^\ast}{m_e} \right) m^{-1}\lambda^{-2} 
  \end{equation}
  in the light of the expression (\ref{Eq-10}) derived here, where $m^\ast$ is the effective mass of an electron in a system. One difficulty in comparison between theory and experiment is that in much of the literature on conventional superconductors, only the \textit{change} of penetration depth with respect to a given temperature rather than the absolute value of $\lambda$ has been measured in bulk sample. Absolute values have been measured for colloidal particles \cite{Shoenberg,Lock} and large area thin films on mica, but for those samples it is difficult to estimate other properties like resistivity and carrier density which could significantly differ from bulk and have not been reported. Nevertheless, researchers used indirect schemes to estimate $\lambda (0)$. For example, in Ref.~\onlinecite{Rinderer83} for Pb, $\Delta$ obtained from tunneling was used as input parameter and $\lambda (0)$ was obtained from tuning it to the value that consistently reproduced the BCS temperature dependence $\lambda (T)$ for a set of samples with different amount of impurity. In some other cases such as in pure Sn crystal \cite{Devlin}, $\lambda (0)$ was estimated from the normal state properties. More recently, absolute measurement of $\lambda$ have been performed on a number of superconducting thin films using two-coil mutual inductance technique \cite{Benfatto,Kamlapure,Bose} and on some single crystals using microwave techniques \cite{Hafner}. 
  Here, we analyze the data of Nb-doped STO \cite{Scheffler} and Sn crystal \cite{Devlin}, polycrystalline \cite{Rinderer83} Pb and 15.3 nm thick Nb film \cite{Lemberger07}, and relatively stronger disordered thin films \cite{Mondal11,Mandal20} of NbN and {\it a}-MoGe.
 % Nb-doped SrTiO$_3$ is believed to be a multi-band superconductor \cite{Bednorz} but this detail is not very important in the present context since the temperature dependence of $\lambda(T)$  can be effectively described by a single superconducting energy gap due to large interband scattering. Together these systems span a large range of disorder for which $n_s/n \sim 0.6$--$10^{-4}$.
 Although Nb-doped SrTiO$_3$ was initially thought to be a  multi-band superconductor \cite{Bednorz}, the recent data are in favor of a single-band \cite{Thiemann2,Hwang,Eagles} superconductor. Together these systems span a large range of disorder for which $n_s/n \sim 0.6$--$10^{-4}$.
  In Table~\ref{Table1}, we summarize the properties of these materials. For Sn and Pb, the authors reported $\lambda (T)$ vs. $\left( 1-(T/T_c)^4\right)^{1/2}$; the data was digitized and converted into $\lambda^{-2}(T)$ vs. $T$. 
  One important parameter in Table~\ref{Table1} is the effective mass of the electron. This value is taken either from electronic specific heat (Sn, Pb, NbN) or quantum oscillations (Nb-doped STO and Nb). For a-MoGe, we did not find an independent estimate but used the electron mass as has been done in the literature \cite{Tashiro}. 
  In figure~3(a)--(g), we show the temperature variation $n_s/n$ for different materials.
   We first focus on the Nb-doped SrTiO$_3$ crystal which is the cleanest sample analyzed here. In Fig.~3(a) we fit $n_s(T)/n$ using the full expression in Eq.(\ref{Eq-10}) using the values of $\delta$ as shown in Table~\ref{Table2} and $\alpha$ as the only adjustable parameter. In the same panel we also separately plot the 1st and 2nd term on the right hand side of Eq.~(\ref{Eq-10}). In Fig. 3(b), we try to fit the same data using only first term which is equivalent to the dirty limit expression in Eq.~(\ref{Eq-0}). As can be seen best fit curve deviates at high temperature, showing  at this level of cleanliness a small but discernible difference in the T-dependence emerges between the exact expression and the dirty-limit BCS expression.
    For Sn, Pb, Nb film (Fig.~3(c)-3(e)) as $n_s/n$ decreases, the contribution of the 2nd term in the overall expression progressively decreases. For the strongly disordered NbN and a-MoGe films (Fig.~3(f)-3(g)) the contribution of the 2nd term is negligible and the data can be fitted with the dirty limit BCS expression.  The extracted parameters from the fits are also shown in Table~\ref{Table2}. Wherever resistivity data is available the values of $\tau$ extracted from the present fits, $\tau_{_P}$ are consistent with those obtained from resistivity, $\tau_{_T}$, using Drude model.  
  %The experimental data is plotted along with their fits with Eq.~(\ref{Eq-10}), using the value of $\delta$ as shown in Table~\ref{Table2} and $\alpha$ as the single adjustable parameter. The extracted parameters from the fits are also shown in Table~\ref{Table2}. The values of $\tau$ obtained from the fits are consistent with that obtained from resistivity using the measured values of carrier density using the free electron formulae. In addition, in panels 3(a)--3(c), we separately plot the temperature dependence of the 1$^{\rm st}$ and 2$^{\rm nd}$ term on the right hand side of Eq.~(\ref{Eq-10}). For the single crystal of Sn, which is the cleanest sample analyzed here the 2$^{\rm nd}$ term is about 10\% of the 1$^{\rm st}$ term. For progressively dirtier systems (smaller $n_s(0)/n$) the second term becomes less important. For the strongly disordered NbN and MoGe, the second term can be completely ignored and the data can be fitted with only the first term which corresponds to the well-known dirty limit BCS expression. 
  In Fig.~3(h), we show the ratio of the second term to the first term, $\eta$, as a function of $\Delta_0\tau$. 
  It is obvious from the graph that the cleanest superconductor analyzed here, Nb-doped STO, is far from the BCS clean limit for which $n_s(0)/n \sim 1$ and $\Delta_0 \tau >>1$.
  Most studies on pure elemental superconductor show $n_s/n = 0.05$--$0.3$ \cite{Tai,Pippard,Dressel}. 
  %Somewhat smaller values of $\lambda(0)$ in a Sn single crystal was reported by Tai et al \cite{Tai} which furthermore was dependent along the crystallographic orientation along which it was measured. If we consider the smallest value of $\lambda(0)$, we obtain $n_s/n ~ 0.3$. Unfortunately, in that paper the data is not presented in a way that we can easily read out the $\lambda(T)$ as a function of temperature.  
  %Even though we have not analyzed the penetration depth of Al here, Faber and Pippard \cite{Pippard} had also obtained $n_s/n \sim 0.1$ from early microwave studies in polycrystalline Al.
  %We see from the incomplete nature of the data spread over several decades, 
  Surprisingly, there is one report \cite{Strongin} where $n_s/n$ values very close to one was reported for very pure polycrystalline Ta and Nb. However, in that paper $\lambda(0)$ values were obtained from $\lambda(T)$ close to $T_c$. However for the same sample, the low temperature variation of $\lambda(T)$ showed unexpected distinct deviation from BCS variation, probably from surface contamination.
  Similarly it was suggested that Nb-doped SrTiO$_3$ could be in the clean limit \cite{Collignon} but this has been contested from direct measurements of the penetration depth \cite{Scheffler}. Therefore there is a need
  for further measurements on high purity single crystals to explore if the BCS limit can indeed be realized.   
  
  % I have not touched the section on comparison between theory and experiment. I do not know what is he origin of the notes on it. I see that it is one long paragraph. For ease to the reader, it should be brojen up into more than one, I think. 

\section{Outlook and Conclusion}

Our analysis is based on the Born approximation for disorder potential. We thus have not considered localization effect which plays a major role for strongly disordered superconductors when $k_F\ell \sim 1$ (where $k_F$ is the Fermi wavenumber and $\ell$ is the mean free path of an electron). The superfluid density presented here is without consideration of higher order effects due to phase fluctuations which again finds its role for relatively large disorder when $\alpha=\Delta_0\tau/\hbar \lesssim 10^{-5}$, and hence the physics of pseudogap phase \cite{Pratap_jpcm} has also been ignored.

Our study reveals that the absolute measurement of superfluid density at all temperatures, rather than the relative measurement with respect to a given $T$, is necessary for determining its dependence on disorder. This is because $n_s(T)/n_s(0)$ is weakly disorder dependent while both $n_s(T)$ and $n_s(0)$ are disorder dependent. 
This analysis is based on the assumption that $\Delta$ is disorder independent, as a consequence of Anderson's theorem \cite{Anderson}.

We find that the estimated relaxation time from the resistivity data and from the fitted parameter $\alpha$ are in the same ballpark for all the samples those have been analyzed, excepting purer samples Pb and Sn for which resistivity data are not available for comparison. 
One surprising finding in this study is that most samples on which the temperature dependence of the superfluid density has been investigated seem to be in the dirty limit where $n_s(0) << n$. In fact, the paradigmatic BCS clean limit seems to be very rare. To achieve the clean BCS limit the superconductor needs to have a large electronic relaxation time, $\tau > \hbar/\Delta_0 \sim 10^{-11}$--$10^{-12}$s, which translates into an electronic mean free path, $\ell$, greater than tens of micrometers. Such a large $\ell$ is indeed very rare and has been realized in very high purity single crystals of noble metals like Ag and semimetals like Bi on which electron focusing experiments \cite{Benistant83,Heil} were performed. This requirement is even more stringent than the mean free path required in typical single crystals on which de Haas-van Alphen measurements are performed at fields of \textit{several Tesla}. 
It will be instructive to try to synthesize superconductors with comparable mean free path to experimentally verify the temperature variation of $n_s/n$
 from the clean-limit BCS theory.

%Whether it is possible to realize comparable mean free path in a single crystal of a superconductor is an outstanding question that needs to be addressed through careful experiments in future. 

\begin{widetext}
	
	\begin{table}[h]
		\caption{Parameters calculated using or extracted from the experimental data shown in table~\ref{Table1} for all the samples. Relaxation time calculated using the transport data, $\tau_{_T} = m^\ast/(ne^2 \rho_{_N})$, and the same calculated using the parameter $\alpha$ extracted by fitting $n_s/n$ with Eq.~(\ref{Eq-10}), $\tau_{_P}$, are in the same ballpark.}
		\begin{tabular}{|l|l|l|l|l|l|l|}\hline
			Sample &  $n_s(0)$  & $\frac{n_s(0)}{n}$ & $  \delta=\frac{\Delta(0)}{2k_BT_c}$ & $\alpha= \frac{\Delta (0) \tau}{\hbar}$ & $ \tau_{_T}=\frac{m^\ast}{ne^2\rho_{_N}}$ & $\tau_{_P} = \frac{\alpha \hbar}{\Delta (0)}$ \\
			&  ($10^{25}$m$^{-3}$) & $(10^{-3})$ & &$(10^{-3})$ & ($10^{-17}$ s) & ($10^{-17}$ s) \\ \hline
			Sn & 1967.17 & 132.92 & 0.865 & 48.5 &  ***  & 5751.8\\ \hline
			Pb & 2015.56 & 152.69 & 1.082& 63 &  *** & 3094 \\ \hline
			Nb (15.3nm) & 279.7 &50.3 & 0.96 & 17.7& 855 & 763.9 \\ \hline
			NbN-1 &21.94 &1.30 & 1.0135 &0.415 &18.4& 10.9\\ \hline
			NbN-2& 8.27 & 0.713 & 1.0135 & 0.228 & 13.8& 8.64 \\ \hline
			NbN-3& 4.89 & 0.416 & 1.0135 & 0.134&12.5 & 5.93 \\ \hline
			MoGe-1 (21 nm) & 10.11 & 0.219 & 1.06 & 0.0694 &5.14 & 3.57 \\ \hline
			MoGe-2 (11nm)& 9.17 &0.199 & 1.116 & 0.0638 &4.7 & 3.36 \\ \hline
			MoGe-3 (4.5nm) & 7.50 & 0.163 & 1.3 & 0.0518 & 5.35 &3.04  \\ \hline
			Nb-doped STO & 6.2 & 563.1 & 0.875 & 500 & $2.5 \times 10^5$ \cite{Scheffler} & $6.3 \times 10^5$ \\ \hline
		\end{tabular}
		\label{Table2}	
	\end{table}

\end{widetext}

\vspace{2cm}

\appendix

\section{Sum Rule for the Suppression of Superfluid Density}

While the clean BCS limit can only be reached in specially prepared very clean single 
 crystals,  frequently available polycrystalline and thin film superconductors are in the opposite limit, i.e., dirty limit where, $\tau \ll \Delta_0/\hbar$. In such a situation, $n_s(0) \ll n$. $n_s/n$ can be intuitively estimated based on the oscillator sum rule \cite{Kubo,Glover,Ferrell} that gets the result correct within a factor of order unity; 
 here we outline this derivation and compare with the accurate expression of $n_s$ that
 has already been derived microscopically in this paper (\ref{Eq-10a}) and originally by Abrikosov and Gorkov \cite{A-G} in the linear response theory.

The optical conductivity of a metal in Drude theory is given by $\sigma (\omega) = \sigma' (\omega) + i \sigma'' (\omega)$ where
\begin{equation}
\sigma' (\omega) = \frac{\sigma_0}{1+ (\omega \tau)^2} \,\,\, ; \,\,\,  \sigma'' (\omega) = \frac{\sigma_0\omega \tau}{1+ (\omega \tau)^2}
\end{equation}
with dc conductivity $\sigma_0 = ne^2\tau/m_e$. The well known oscillator sum rule for $\sigma'(\omega)$ is given by
\begin{equation}
\int_0^\infty \sigma'(\omega) \,d\omega =\frac{\pi n e^2}{2m_e} \, . \label{SR2}
\end{equation} 
The sum rule in Eq.~(\ref{SR2}), however, remains unaltered for finite temperature, magnetic field, the presence of interaction between electrons, and even when the metallic system makes a phase transition into the superconducting state. However, the spectral weight in $\sigma'(\omega)$ is redistributed, depending on the state of the system.

When a metal goes into the superconducting state, a spectral gap opens for the frequency
$\omega<2\Delta_0/\hbar$. At a very high frequency $(\omega >> 2\Delta_0/\hbar)$, the distribution of spectral weight in the real part of conductivity in the superconducting state, $\sigma_s'(\omega)$, remains unaltered from its metallic counterpart. $\sigma_s'(\omega)$ approaches zero as $\omega \to   2\Delta_0/\hbar$ from its higher values. However, this depletion of spectral weight gets accumulated at zero frequency in the form of Dirac delta function:
\begin{equation}
\sigma'_s(\omega) = \frac{\pi n_s e^2}{m_e}\delta(\omega)
\end{equation}
where the prefactor $\pi n_s e^2/m_e$ is known as Drude weight to the conductivity that is proportional to the superfluid density.
The precise variation of $\sigma_s(\omega)$ for a s-wave superconductor may be obtained from
Mattis-Bardeen theory\cite{Bardeen}. However for the purpose of an approximate estimation of $n_s$, we consider a discontinuous jump in $\sigma'_s(\omega)$ at $\omega = 2\Delta_0/\hbar$ from its zero value to normal-metallic value. Following the sum rule (\ref{SR2}), we thus write
\begin{equation}
\int_0^{2\Delta_0/\hbar} \sigma'(\omega) d\omega \approx \int_0^{2\Delta_0/\hbar} \sigma'_s(\omega) d\omega
\end{equation} 
which yields
\begin{equation}
\frac{n_s}{n} = \frac{2}{\pi} \tan^{-1} \left( \frac{2\Delta_0\tau}{\hbar} \right) 
\label{SR4}
\end{equation}
reproducing the clean limit ($\Delta_0 \tau \to \infty$), i.e., $n_s =n$. In the dirty limit ( $\Delta_0 \tau \to 0$), we find $n_s/n = 4\Delta_0\tau/ (\pi \hbar)$ which differs with the microscopic result only by a numerical factor $\pi^2/4$. 

It is instructive to write Eq.(\ref{SR4}) in terms of the measurable quantities such as penetration depth and normal state resistivity $\rho_{_N} = 1/\sigma_0$. Substituting $n_s$ by $(m_e/\mu_0e^2) \lambda^{-2}(0)$ in  Eq.(\ref{SR4}) and reinstating the above mentioned factor $\pi^2/4$, we find 
\begin{equation}
\lambda^{-2}(0) = \frac{\pi \mu_0\Delta_0}{\hbar \rho_{_N}}
\label{SR6}
\end{equation}
in the dirty limit. 
The relation (\ref{SR6}) is particularly powerful as it relates three independent measurable quantities $\lambda(0)$, $\Delta_0$ and $\rho_{_N}$ without any adjustable parameters.

\section*{Acknowledgments}
 
 PR would like to thank Mohit Randeria for valuable discussions in 2008 on the connection between the oscillator sum rule and superfluid density. We thank Thomas Lemberger and Marc Scheffler for sharing data on Nb and Nb-doped SrTiO$_3$ respectively. We also thank Marc Scheffler and Peter Armitage for valuable online discussions and feedback after an early draft of this paper was circulated.
   We acknowledge financial support by the Department of Atomic Energy, Govt of India (Project No: RTI4003).

\end{document}